\documentclass{PoS}

\usepackage{amsmath}
\usepackage{xspace}

\newcommand{\particle}[1]{{\ensuremath{#1}}\xspace}

\newcommand{\Bs}{\particle{B^0_s}}
\newcommand{\Bd}{\particle{B^0}}
\newcommand{\Bu}{\particle{B^+}}
\newcommand{\mup}{\particle{\mu^+}}
\newcommand{\mum}{\particle{\mu^-}}
\newcommand{\Jpsi}{\particle{J\!/\!\psi}}
\newcommand{\Jmm}{\particle{\Jpsi(\mu\mu)}}

\newcommand{\decay}[2]{\particle{#1\!\to #2}\xspace}
\newcommand{\Bsmumu}{\decay{\Bs}{\mup\mum}}
\newcommand{\BsKK}{\decay{\Bs}{ K^+K^-}}
\newcommand{\BdKpi}{\decay{\Bd}{ K^+\pi^-}}
\newcommand{\Jpsimumu}{\decay{\Jpsi}{ \mu^+\mu^-}}
\newcommand{\BuJpsimumuK}{\decay{B^+}{\Jmm K^+}}

\newcommand{\unit}[1]{\ensuremath{\,{\rm #1}}\xspace}

\newcommand{\tev}{\unit{TeV}}

\newcommand{\mevcc}{\unit{MeV\!/\!{\it c}^2}}

\newcommand{\invfb}{\unit{fb^{-1}}}

\title{The LHCb analysis for {\boldmath \Bsmumu } }

\ShortTitle{The LHCb analysis for \Bsmumu}

\author{\speaker{Marc-Olivier Bettler}\thanks{The material presented at the FPCP conference is available on the CERN document server~\cite{poster}.}\\
        on behalf of the LHCb collaboration\\
        Ecole Polytechnique F\'ed\'erale de Lausanne (EPFL)\\
        E-mail: \email{marc-olivier.bettler@cern.ch}}

\abstract{LHCb, bolstered up by the $10^{12}$ $b$-hadrons produced yearly, is an excellent place to study rare $B$ decays. 
The \Bsmumu decay, generated by flavour-changing neutral current, is strongly suppressed within the Standard Model (SM). 
However, its branching ratio can be significantly enhanced if New Physics (NP) exists. 
The current best limit is still an order of magnitude above the SM prediction, thus leaving room for observation of NP effects. 
In this document, the current LHCb analysis strategy for the \Bsmumu search is presented.
}

\FullConference{Flavor Physics and CP Violation 2009\\
		 May 27-June 1 2009\\
		 Lake Placid, NY, USA }

\begin{document}

\section{Introduction}

The Standard Model (SM) is successful in explaining almost all observations in particle physics experiments so far. 
Nevertheless, there are reasons to consider it as a low energy effective limit of a more general theory. 
In that prospect, observables for processes where the SM contribution is highly suppressed are particularly interesting. 
Within the SM, flavour-changing neutral current processes are highly suppressed since they are forbidden at tree level and can only proceed via loops diagrams. 
If New Physics (NP) exists, new particles can contribute to those processes in the loop diagrams, modifying observable quantities with respect to the SM prediction. 
The \Bsmumu decay branching fraction is  such an observable.
LHCb~\cite{LHCb} will take advantage of the copious $b$-hadron expected at LHC~\cite{LHC}. 
LHCb is a single-arm forward spectrometer primarily optimized to the study of CP-violation and rare decays in $b$-hadrons. 
The detector is characterised by its precise vertex detector, powerful particle identification capabilities and versatile trigger. 
Nominally, LHCb will operate at a luminosity $\mathcal{L} = 2 \times 10^{32}$ \unit{cm^{-2}} \unit{s^{-1}}, giving 2 \invfb per year ($10^7$ seconds) of data. 
The analyses presented in this document are applied to Monte Carlo simulated data with a full detector response, including pile-up (multiple $pp$ collisions in a single bunch-crossing) and spill-over (signal coming from particles produced in a previous bunch-crossing).

\begin{figure}[ht]
\begin{minipage}[b]{0.45\linewidth}
\centering
\includegraphics[width=\linewidth]{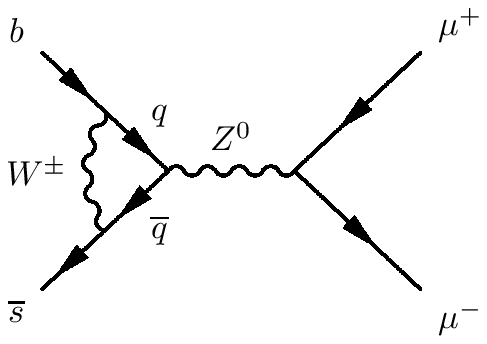}

\caption{\label{feynman_SM}Feynman diagram of the \mbox{\Bsmumu} decay within the SM.}
\end{minipage}
\hspace{0.5cm}
\begin{minipage}[b]{0.45\linewidth}
\centering
\includegraphics[width=\linewidth]{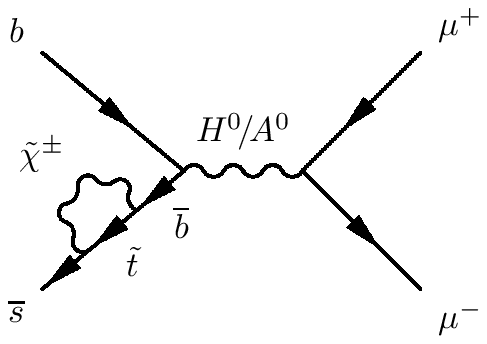}

\caption{\label{feynman_MSSM}Feynman diagram for a MSSM contribution to \mbox{\Bsmumu}.}
\end{minipage}
\end{figure}

Within the SM, the \Bsmumu decay occurs through loops diagrams like the one in Figure~\ref{feynman_SM} and its branching ratio is expected to be ${\rm BR}(\Bsmumu) = ( 3.35 \pm 0.32)\times 10^{-9}$~\cite{SM}. 
In minimal supersymmetric extensions of the SM (MSSM), this decay would receive additional contributions from diagrams of the kind shown in Figure~\ref{feynman_MSSM}. 
The branching ratio is then proportional to the sixth power of the ratio of the Higgs vacuum expectation values, $\tan \beta\xspace$, and can be considerably enhanced~\cite{NUHM}.
The current upper limit given by the two Tevatron experiments is $4.7 \times 10^{-8}$ at $90\%$ CL~\cite{mumu-cdf,mumu-d0}.


\newpage

\section{Strategy for the {\boldmath \Bsmumu} Decay Search}

The current LHCb strategy for the \Bsmumu decay search~\cite{mumu-1} can be summarized as follows:

\begin{itemize}

\item The analysis relies on three independent variables, related to  the dimuon invariant mass, the  particle identification of the daughters, and  geometrical information from the decay topology, respectively. 
Since the three variables are uncorrelated, they can be calibrated independently. 
The calibration methods are designed to rely solely on real data.

\item For each event, likelihood values are calculated for the each of three variables, under the signal and background hypothesis.

\item The compatibility of the obtained distributions of the likelihood values is tested against \mbox{\Bsmumu} branching ratio hypotheses, using the \emph{CL} modified frequentist method~\cite{Cls} with the calibrated likelihood distributions for signal and background. 
The final result is either a measurement or an upper limit of the branching fraction.

\item Since the number of \Bs produced is not precisely known, the use of a normalisation channel with a well known branching fraction is required to obtain an absolute measurement of the branching fraction or upper limit.

\end{itemize}

\subsection{Invariant Mass Likelihood}

\begin{figure}[ht]
\centering
\includegraphics[width=0.5\linewidth]{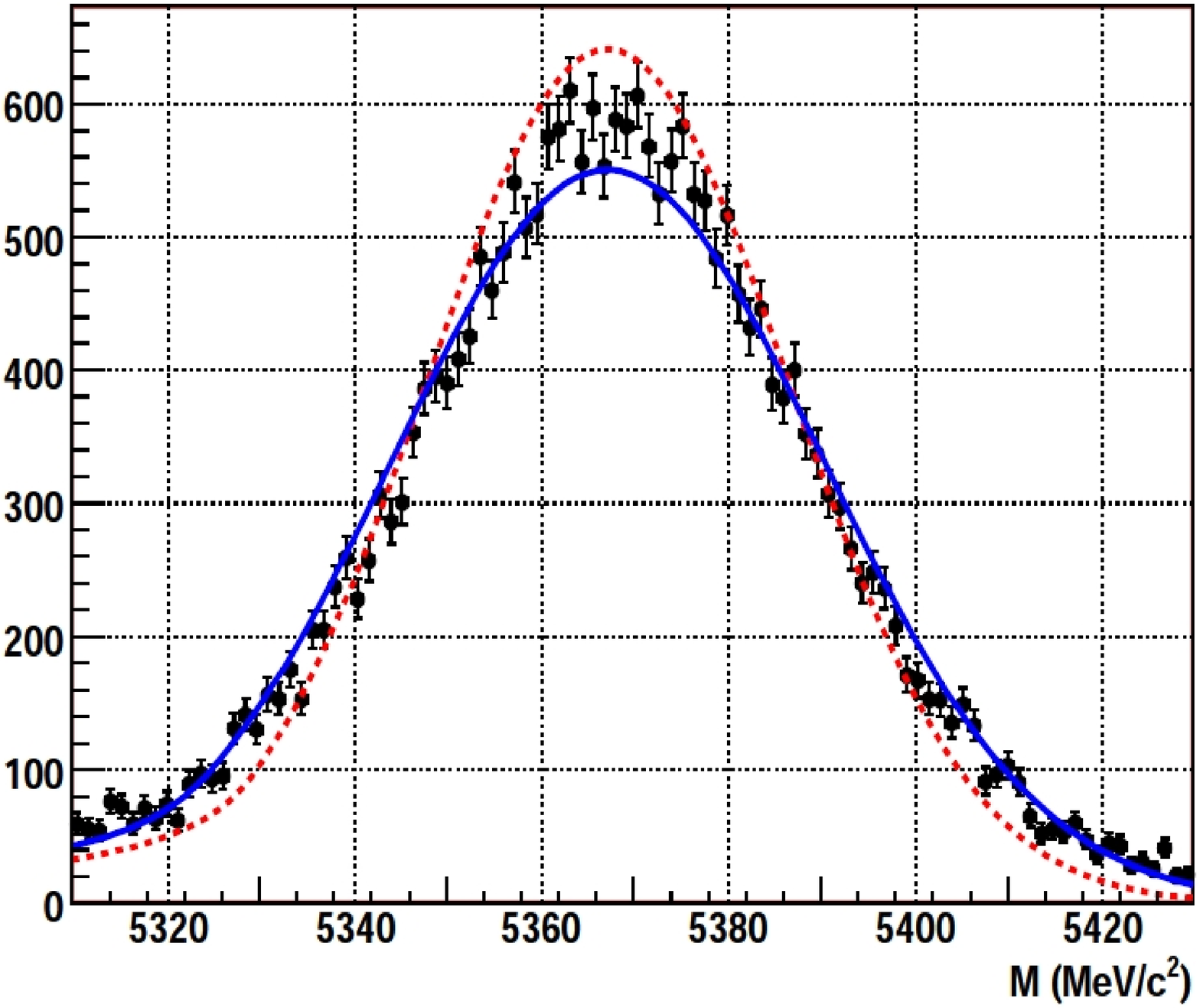}
\caption{\label{mass_sig}Invariant mass distributions for \mbox{\BsKK} with a correction for the kaon identification cut effect (solid curve) and without correction (dashed curve). 
Points with error bars represent \Bsmumu invariant mass distribution.}
\end{figure}

The signal likelihood distribution is calibrated using the \mbox{\BsKK} decay, which is kinematically very close to the signal.
The selection of the \mbox{\BsKK} events uses a cut on the kaon identification, which biases the \Bs mass resolution.
The effect of the kaon selection cut on the mass resolution is assessed on \mbox{\decay{\particle{B^0_{(s)}}}{\particle{h^+h^-}}} control channels, and the obtained correction factor is applied to \mbox{\BsKK} sample (see Figure~\ref{mass_sig}).

The background likelihood is calibrated using \mbox{\Bsmumu} candidates in the mass sidebands, outside the $\pm 60 \mevcc$ region around the nominal \Bs mass.

\subsection{Particle Identification Likelihood}

The identification of the daughter particles as muons relies mainly on the muon system information.
The signal likelihood is calibrated using a high-purity muon sample obtained from \mbox{\Jpsimumu} decays where only one muon is identified.

The background likelihood due to misidentified hadrons is obtained from \mbox{\decay{\Lambda}{$p$ \ \pi^-}} decays, which are very abundant in LHCb and have a very clean signature. 





\newpage

\subsection{Geometrical Likelihood}

The geometrical likelihood is given by a single real number computed from the following topological variables:

\begin{itemize}
\item the \Bs impact parameter with respect to the primary vertex,
\item the \Bs proper time,
\item the smallest impact parameter significance of the two muons with respect to any primary vertex, 
\item the distance of closest approach between the two muons,
\item the isolation of each muon track.
\end{itemize}

\begin{figure}[ht]
\begin{minipage}[t]{0.5\linewidth}
\centering
\includegraphics[width=\linewidth]{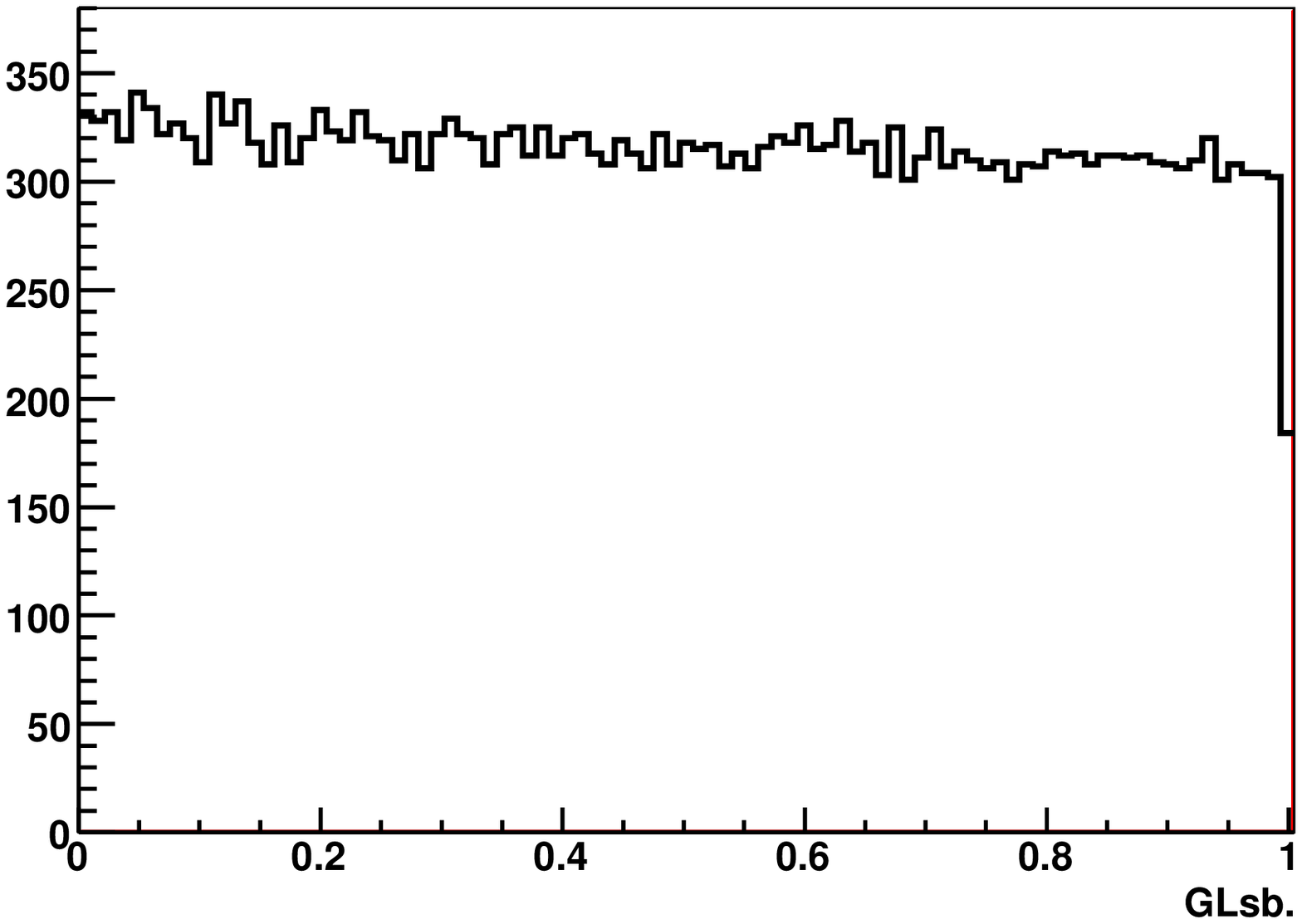}
\caption{\label{GL_sig}Geometrical likelihood distribution of the signal obtained by simulation.}
\end{minipage}
\hspace{0.5cm}
\begin{minipage}[t]{0.5\linewidth}
\centering
\includegraphics[width=\linewidth]{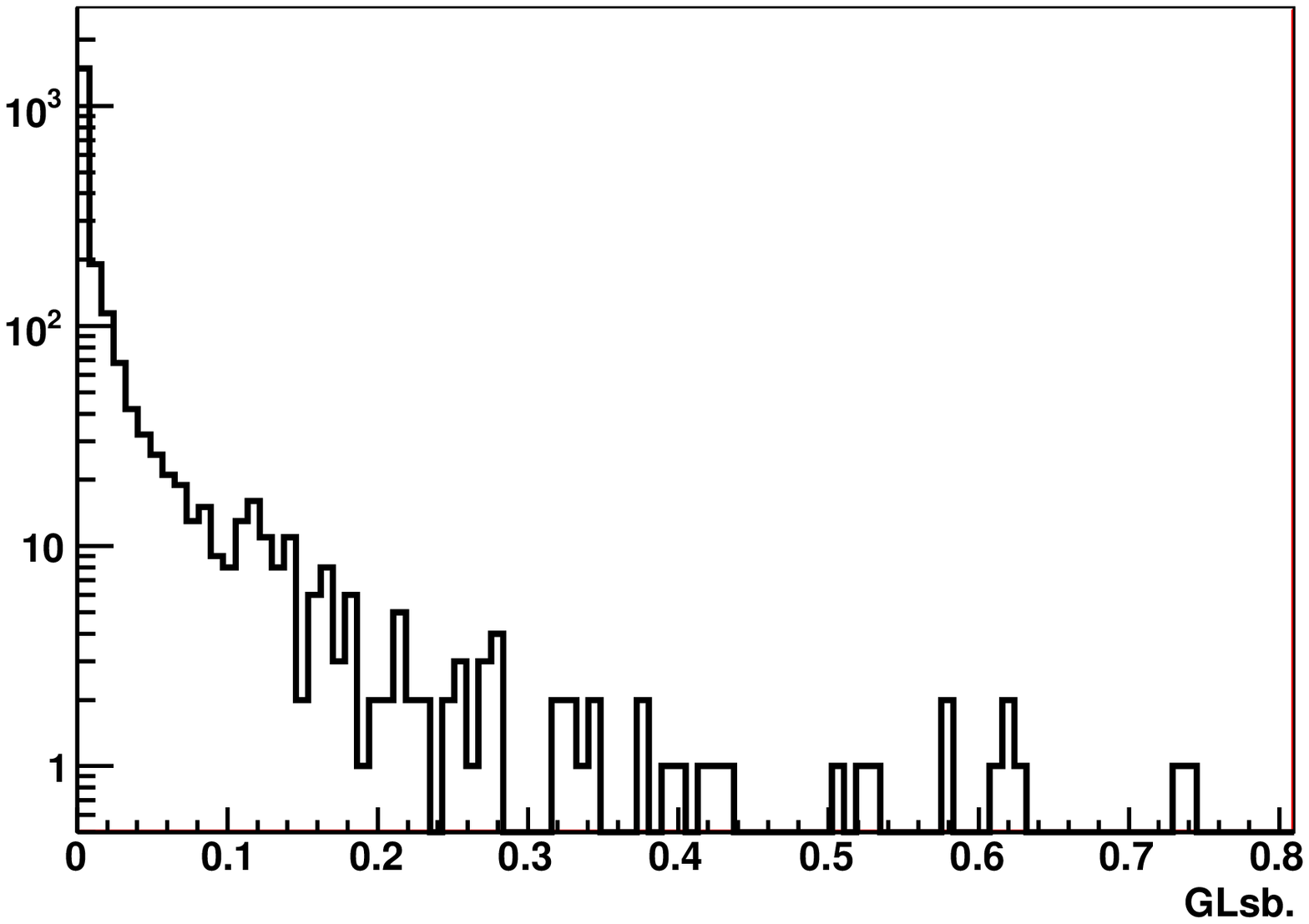}
\caption{\label{GL_bkg}Geometrical likelihood distribution of background events containing two muons in the detector acceptance.}
\end{minipage}
\end{figure}

Using Monte Carlo samples, this likelihood is defined to take values between 0 and 1, and to have a flat distribution for signal events and a distribution peaked at 0 for background events, as shown in Figures \ref{GL_sig} and \ref{GL_bkg}. 

The signal distribution will be calibrated by real data using \mbox{\decay{\particle{B^0_{(s)}}}{\particle{h^+h^-}}} events, which are topologically identical to \Bsmumu and the background  distribution by events in the \Bs mass sidebands.

\section{Normalisation}

The branching fraction is expressed as:

\[
 {\rm BR}(\Bsmumu) = \frac{ \mathcal{N}_{\rm sig} }{ 2 \ \sigma_{b\overline{b}} \ \mathcal{L}_{\rm int}  \times \ f_s \ \times \ \epsilon_{ \rm sig } } ,
\]

\noindent where  $\mathcal{N}_{\rm sig}$ is the number of signal events, $\sigma_{b\overline{b}}$ is the $b\overline{b}$ production cross-section, $\mathcal{L}_{\rm int}$ the integrated luminosity,  $ f_s$ the probability of a $b$ quark to hadronize in a \Bs meson, and $\epsilon_{\rm sig}$ the product of the reconstruction, trigger, and selection efficiencies. 
Since $\sigma_{b\overline{b}} \ \mathcal{L}_{\rm int}$ will not be precisely known, we measure the \Bsmumu branching fraction using a normalization channel whose branching fraction is well known.
The \mbox{\Bsmumu} branching fraction is then:

\[
{\rm BR}(\Bsmumu) ={\rm BR}_{\rm norm} \times \frac{f_{\rm norm}}{f_s} \times  \frac{\epsilon_{\rm norm}}{\epsilon_{\rm sig}} \times \frac{ \mathcal{N}_{\rm sig} }{\mathcal{N}_{\rm norm}} ,
\]
\noindent where $f_{\rm norm}$, $\epsilon_{\rm norm}$ and $\mathcal{N}_{\rm norm}$ are the quantities for the normalization channel, with definitions analogous to those of the signal channel. 

Possible normalisation channels are \BuJpsimumuK and \BdKpi.
Particular care has been taken to analyze the signal and normalization channels in a common way so that any large systematic effects in the efficiency ratio cancel. 

The main systematic arises from the $\sim 13 \%$ uncertainty on the ratio $\frac{f_{\Bd}}{f_s}$ or $\frac{f_{\Bu}}{f_s}$. 
With sufficient statistics, all other systematic uncertainties are expected to get much smaller than that, as the analysis relies solely on data.
In the future, pending on a more precise measurement of its branching fraction at Belle, \decay{\Bs}{D^-_s \pi^+} can be used as normalization channel, with the effect of eliminating the use of the factor $\frac{f_{\rm norm}}{f_s}$.

\section{Analysis Sensitivity}

\begin{figure}[ht]
\begin{minipage}[b]{0.48\linewidth}
\centering
\includegraphics[width=\linewidth]{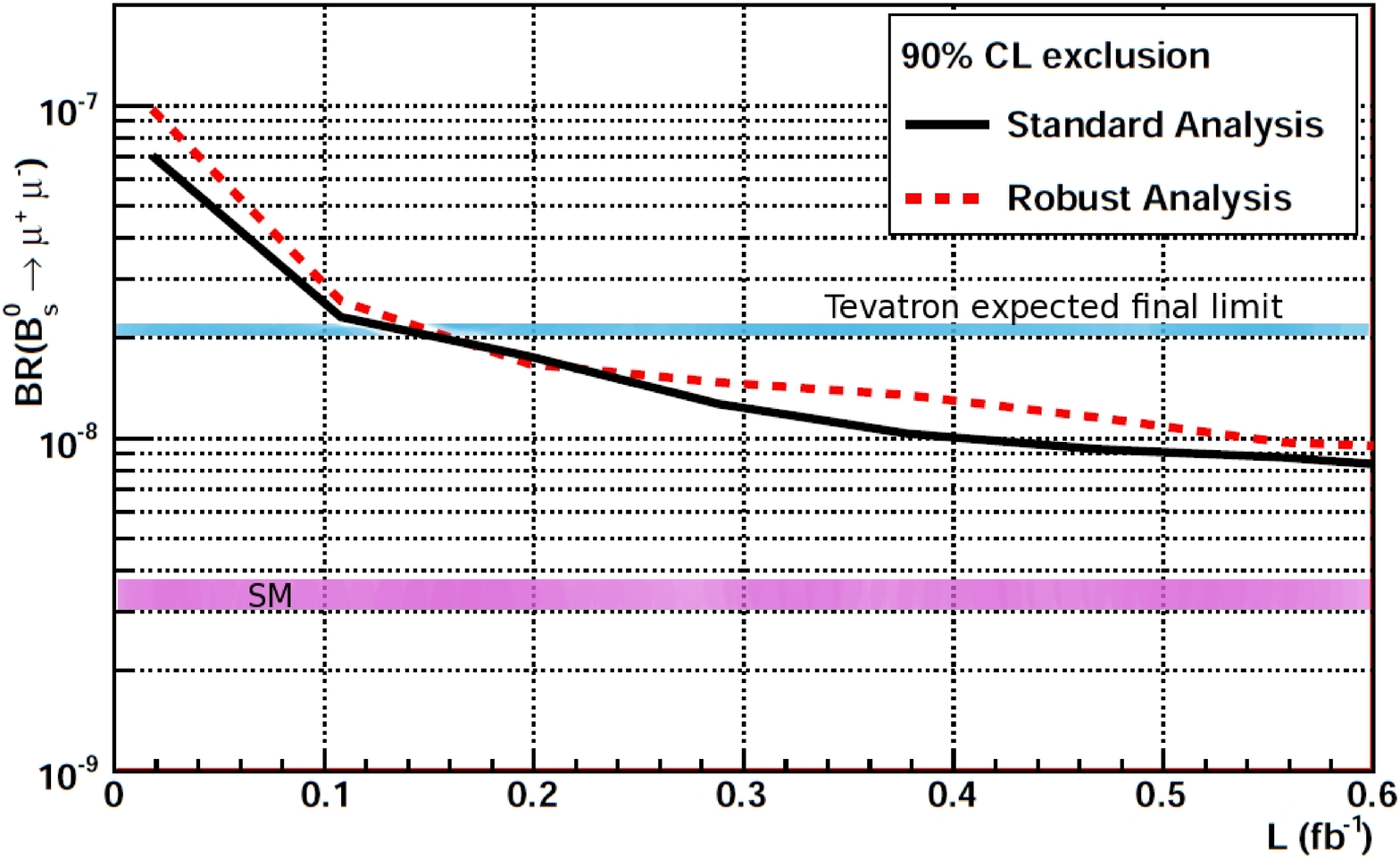}
\end{minipage}
\begin{minipage}[b]{0.52\linewidth}
\centering
\includegraphics[width=\linewidth]{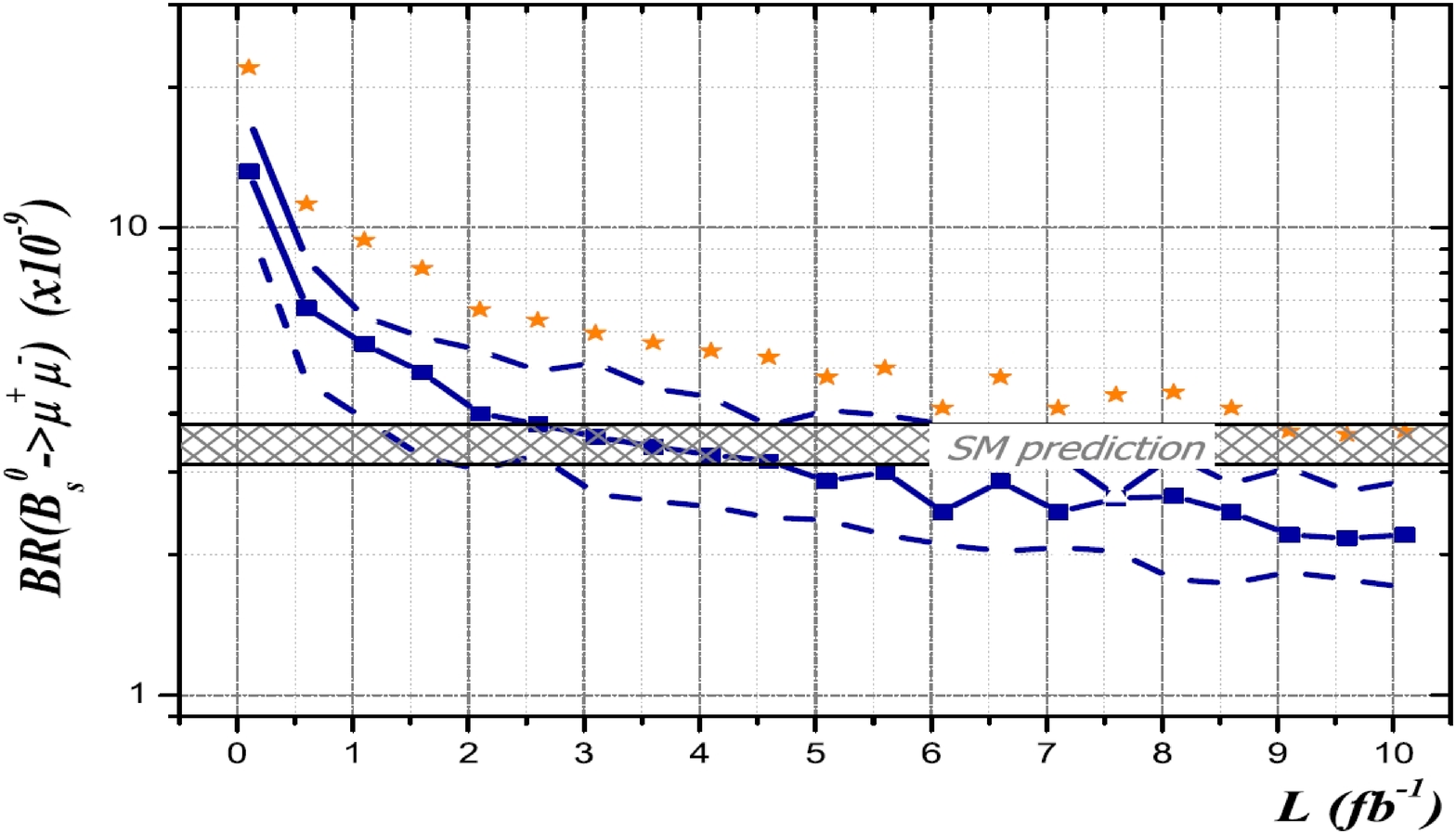}
\end{minipage}
\caption{\label{sensitivity}Expected $90\%$ CL upper limit on {\rm BR}(\Bsmumu) in absence of signal for 8 \tev collisions (left)  and {\rm BR}(\Bsmumu) at which a $5\sigma$ discovery, stars, or a $3\sigma$ evidence is expected, plain blue curve, for 14 \tev collisions (right) as a function of the integrated luminosity.
On the exclusion plot, the black curve is the result of the standard analysis, and the red dashed curve is the result of the robust analysis.
The background estimate has conservatively been set to its  $90\%$ CL upper limit in the exclusion plot, and similarly the dashed curves indicates the 90\% CL upper and lower limit in the observation case. 
The Tevatron limit is calculated by extrapolating the current result to 8 \invfb per experiment. 
}
\end{figure}

The expected sensitivity of LHCb to the \Bsmumu decay as a function of the integrated luminosity is shown in Figure~\ref{sensitivity}.
The solid line in the left plot shows the expected upper limit, at 90\% confidence level, on the branching fraction when no signal is observed, for $pp$ collisions at $\sqrt{s}= 8$ \tev.
The branching fraction for which a $5\sigma$ discovery or for which  a $3\sigma$ evidence is expected, is shown on the right plot for 14 \tev collisions.

The left plot also indicates the expected final limit from the Tevatron experiments, extrapolating the current results to 8 \invfb of data per experiment.
It shows that LHCb competes with the current Tevatron limit ($4.7\times10^{-8}$) with approximately  0.1 \invfb of data and overtakes the final expected limit with about 0.2 \invfb. 
NP models with high $tan \beta$ value are strongly constrained in the process.
In that time frame, the detector may not be fully understood yet. 
Therefore, an alternative robust analysis is designed using variables with a similar physical content to the ones of the standard analysis, but avoiding the use of error estimates. 
This implies a modified selection and definition of the geometrical likelihood.
The robust analysis sensitivity is also depicted in Figure~\ref{sensitivity} (left), as a red dashed curve.
The robust analysis presents  a sensitivity compatible with the one of the standard analysis. 
Therefore, it constitutes a valuable option for the early data. 


About 3 \invfb are enough for a $3\sigma$ evidence if the branching fraction is the SM prediction.
Any enhancement driven by NP will be observed sooner.
Particularly, if the branching fraction is as high as $2\times10^{-8}$, as predicted in Reference~\cite{NUHM}, a $5\sigma$ discovery is possible with very little luminosity ($< 0.4$ \invfb).
With 10 \invfb, a $5\sigma$ discovery occurs if the branching fraction is at the level of the SM prediction.

\end{document}